\documentclass{article}
\newcommand{\documentdate}{17 1 2014}
\usepackage{a4wide,graphicx,color}

\usepackage{amssymb}
\usepackage{amsbsy}
\usepackage{rotating}
\usepackage{enumerate}
\usepackage{amsmath}
\usepackage{rotating}
\usepackage{amsthm,mathrsfs,amsopn}

\usepackage{graphicx}

\usepackage{chemarrow}

\title{Stochastic patterns in a 1D Rock--Paper--Scissor model with mutation}
\author{Claudia Cianci and Timoteo Carletti}
\date{\documentdate}

\begin{document}
\begin{titlepage}

\includegraphics[height=3.5cm]{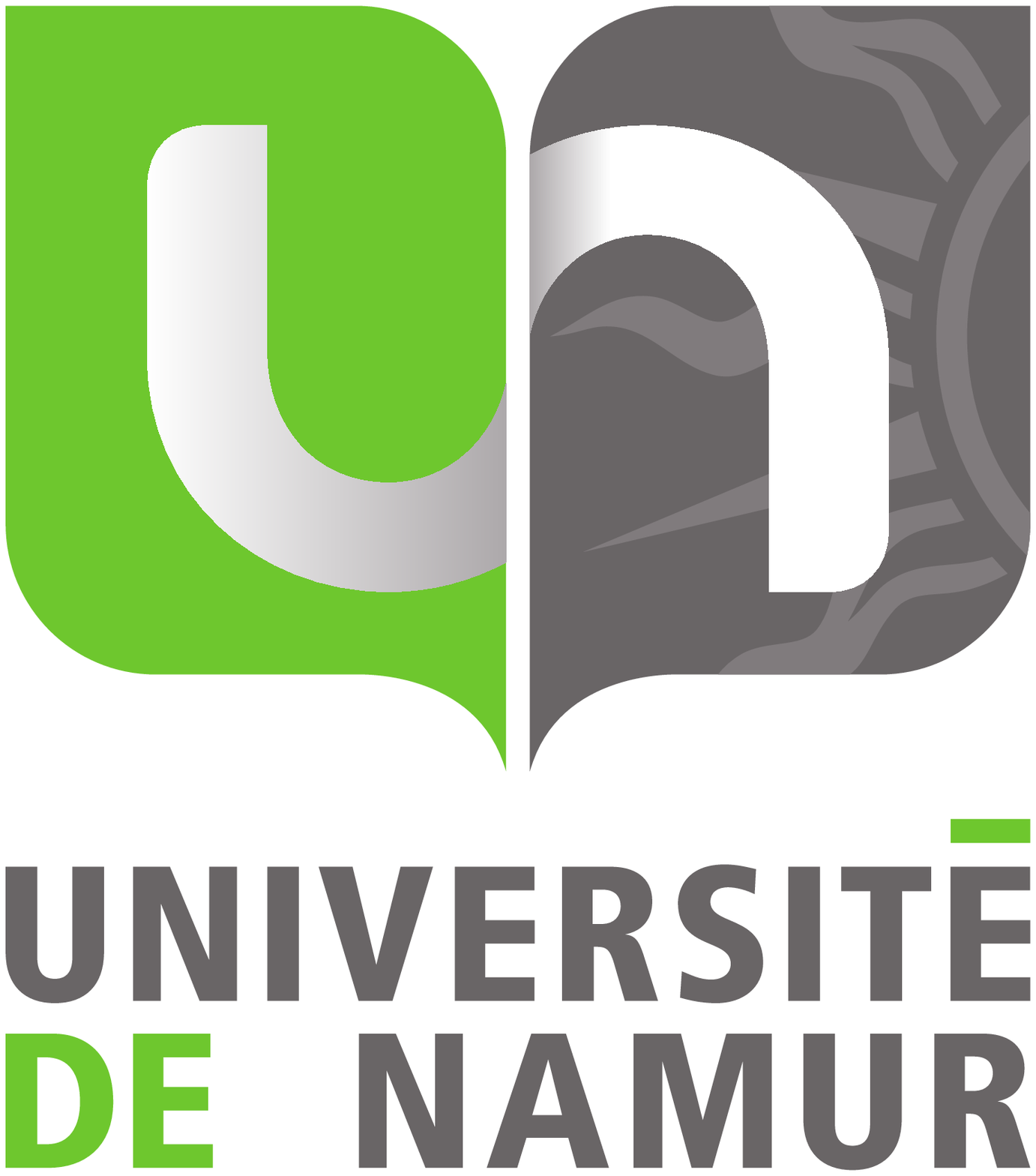}\hfill\includegraphics[height=2.5cm]{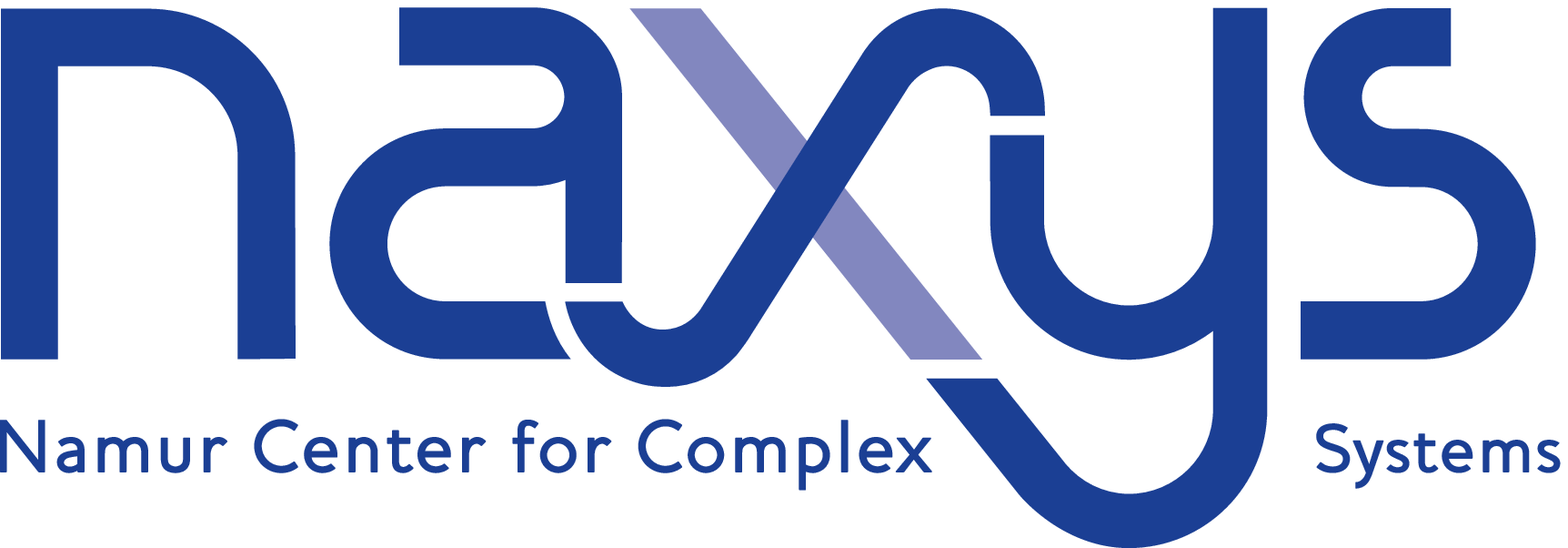}

\vspace*{2cm}
\hspace*{1.3cm}
\fbox{\rule[-3cm]{0cm}{6cm}\begin{minipage}[c]{12cm}
\begin{center}
\vspace{1cm}
Stochastic patterns in a 1D Rock--Paper--Scissor model with mutation\\
\vspace{1cm}
by Claudia Cianci and Timoteo Carletti \\
\mbox{}\\
Report naXys-1-2014 \hspace*{20mm} \documentdate \\
\vspace{2cm}
\includegraphics[width=0.4\textwidth]{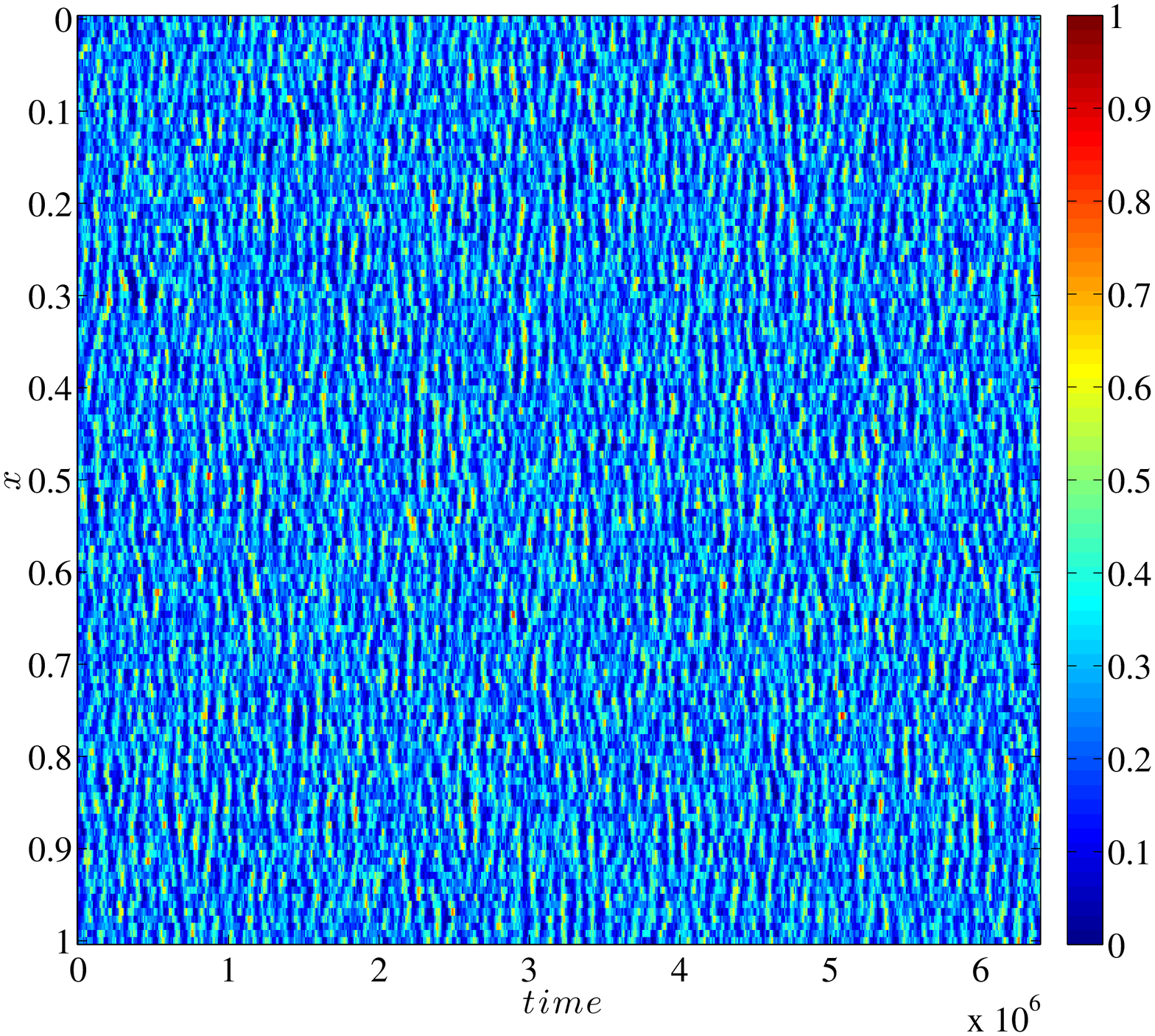}\quad\includegraphics[width=0.5\textwidth]{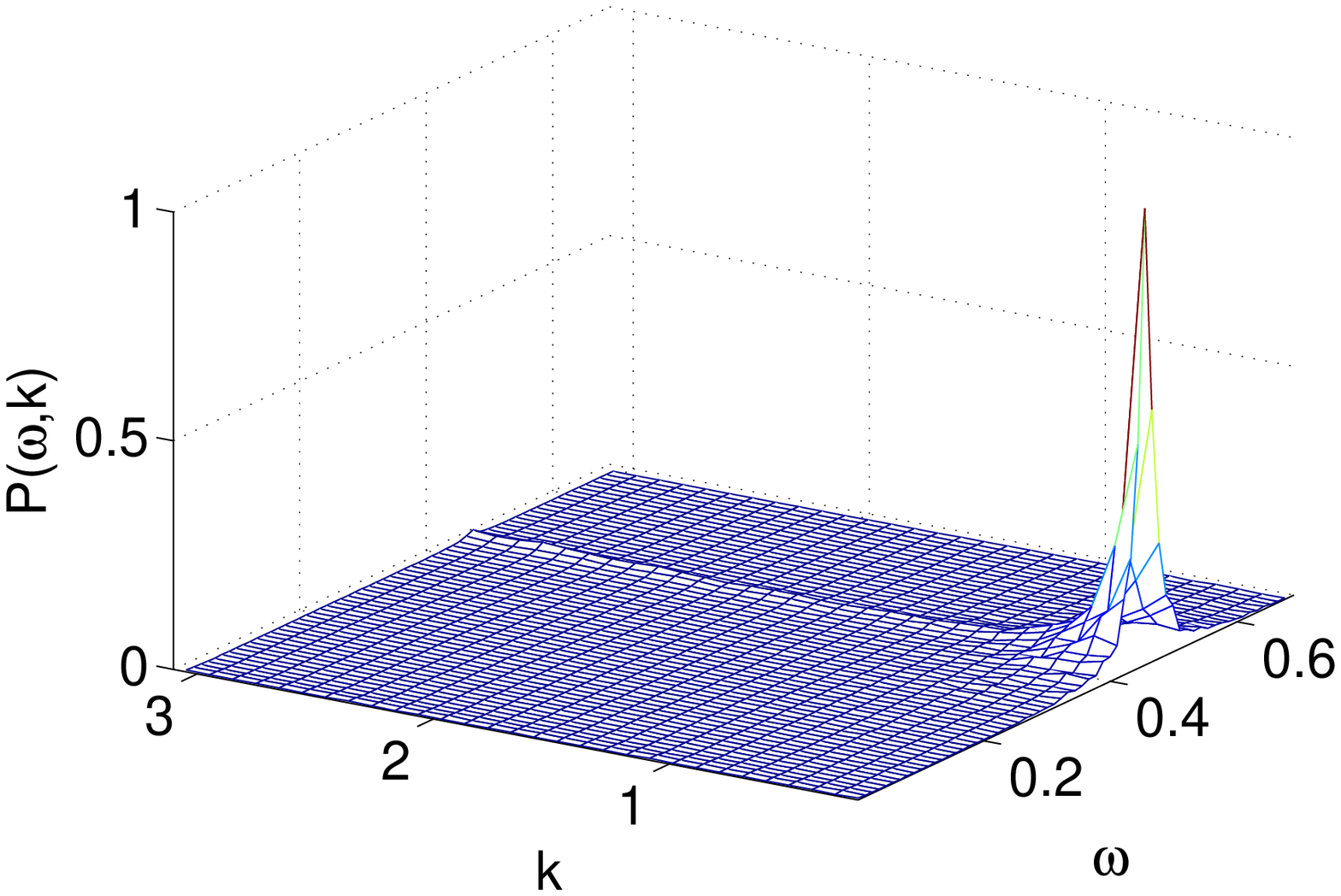}
\mbox{}
\end{center}
\end{minipage}
}

\vspace{2cm}
\begin{center}
{\Large \bf Namur Center for Complex Systems}

{\large
University of Namur\\
8, rempart de la vierge, B5000 Namur (Belgium)\\*[2ex]
{\tt http://www.naxys.be}}

\end{center}

\end{titlepage}
\newpage

\maketitle

%\includegraphics
%\usepackage[italian]{babel}
%\usepackage[italian]{babel}
%\usepackage[T1]{fontenc}\mathbf

%\title{Stochastic patterns in a 1D Rock--Paper--Scissor model with mutation}
%
%\author{Claudia Cianci$^{1}$, Timoteo Carletti$^{2}$}
%
%\affiliation{1. Dipartimento di Sistemi e Informatica and INFN, University of Florence, Via S. Marta 3, 50139 Florence, Italy\\
%2. naXys, Namur Center for Complex Systems, University of Namur, rempart de la Vierge 8, B 5000 Namur, Belgium}

\begin{abstract}
In the framework of a 1D cyclic competition model, the Rock--Paper--Scissor model, where bacteria are allowed to mutate and move in space, we study the formation of stochastic patterns, where all the bacteria species do coexist. We modelled the problem using an individual--based setting and using the system size van Kampen expansion to deal with the Master Equation, we have been able to characterise the spatio--temporal patterns using the power spectrum of the fluctuations. We proved that such patterns are robust against the intrinsic noise and they can be found for parameters values beyond the ones fixed by the deterministic approach. We complement such analytical results with numerical simulations based on the Gillespie's algorithm.
\end{abstract}

\maketitle

\vspace{0.8cm}

{\it keywords}: Stochastic processes, Nonlinear dynamics, Spatio-temporal patterns, Stochastic patterns, Stochastic simulations

\vspace{0.8cm}
\section*{Introduction}

Since the pioneering work of A. Turing~\cite{turing1952}, it is well known that {\em spatio-temporal self-organised patterns} can spontaneously emerge in a reaction-diffusion system: a small perturbation of a homogeneous stable equilibrium can be amplified, through the presence of the diffusion, and eventually drive the system into a non homogeneous spatial solution. Such Turing patterns are widespread and investigated, because of their relevance in applications, for instance in chemistry~\cite{Belousov, Strogatz} and biology~\cite{Murray2}. 

In the classical setting, the emergence of Turing instability needs two diffusing interacting species, the activator and the inhibitor one; systems of three~\cite{satnoianu} simultaneously diffusing species have been also considered and shown to display a rich zoology of possible patterns and instabilities. Patterns can also develop if only one species is allowed to diffuse in the embedding medium~\cite{ermentrout}.  Besides such deterministic models where the concentrations of the interacting species are assumed to take continuous values, one can develop an alternative {\em stochastic individual based} description, that accounts for the discrete nature of the involved species and where the stochastic contributions arise from the finite size corrections.

In a series of recent publications, it has been shown that the intrinsic noise is able create stochastic patterns for parameters values for which the deterministic dynamics predicts a stable homogeneous state; the stochastic effects can be amplified via a resonant mechanism and thus give rise to {\em stochastic Turing patterns}~\cite{Goldenfeld, bia,Fanelli_2011,MainiWoolley, cianci_DNC, emma}. 

There is now a well established analytical framework where such stochastic Turing patterns can be studied, that is the van Kampen system size expansion. This method allows us to expand the master equation in the system size; at the first order one recovers the deterministic mean--field model while at the second order, one can obtain a Fokker-Planck equation describing the stochastic fluctuations. Instead of solving explicitly such Fokker-Planck equation, one can infer the presence of the stochastic patterns by studying the power spectrum of the fluctuations. Such analytical results can be complemented by numerical simulations based on the Gillespie's algorithm.

As we will explain later on, the patters we found would not be strictly speaking due to a Turing mechanism requiring than some eigenvalues should change the sign of their real parts  and thus destabilise the homogeneous solution, they will be purely stochastic ones; nevertheless the Fokker-Plank equation would still provide the right framework where to analyse such patterns. Other studies available in the literature exhibits such stochastic patterns, see for instance \cite{MayLeonard, Barkley, Barkley1, FHN}.

The aim of this paper is to study, along the above lines, the existence of stochastic patterns in a Rock--Paper--Scissor model with mutation, where three species diffuse and interact. Such model has been introduced firstly in~\cite{MayLeonard} and more recently generalised by~\cite{Mobilia, Mobilia1}, allowing for mutation and spatial diffusion. In the latter papers, authors proved the existence of spiral waves, in both the deterministic and stochastic model, for small enough mutation rate and characterised the temporal behaviour of such spirals in term of the Hopf frequency of the limit cycle present in the aspatial model. 

For a sake of clarity we hereby restrict our analysis to a 1D spatial model, however our findings could be straightforwardly extended to the 2D case as well. Because the spatio--temporal spectrum of the patterns we found, is very close to the one determined in~\cite{Mobilia, Mobilia1}, we decided to name such patterns {\em1D spiral waves}. Our main result will be to prove the existence of stochastic spiral waves for parameters values beyond the ones provided for the mean field approximation by~\cite{Mobilia, Mobilia1}. Let us finally stress that the hereby proposed approach is different from the Complex Ginzburg--Landau equation used~\cite{frey,Mobilia3} and the multi--scale method used in~\cite{Mobilia1}. To conclude we also performed dedicated stochastic simulations using the Gillespie's algorithm and confirm a posteriori the adequacy of the  predictions obtained from the theoretical power spectrum.

The paper is organized as follows. In section~\ref{S1} we will introduce the model that will be studied in the next section \ref{sec:ME} using the Master Equation and the van Kampen system size expansion and then in the mean--field approximation in section \ref{ssec:timeevolav}. Finally, the section \ref{S3} will be devoted to the derivation of the Fokker--Planck equation and of its use to study the intrinsic stochastic fluctuations.

\section{The Model}
\label{S1}
For a sake of completeness let us briefly present the Rock--Paper--Scissor model with mutation; we refer the interested reader to~\cite{Mobilia, Mobilia1} for a more complete description. Three populations of agents, say bacteria, hereby named $A$, $B$ and $C$ are considered; each bacterium can move, reproduce itself and interact-fight with bacteria of the other species. The competition is metaphorically described by a Rock--Paper--Scissors game, RPS for short, that is, $A$ overcomes $B$, $B$ overcome $C$ that in turn overcomes $A$. 

The non-spatial RPS model possesses an unstable coexistence equilibrium and three unstable equilibria where only one specie survives, while the generic orbit accumulates to a heteroclinic cycle, that is for longer and longer interval of times the amount of two populations of bacteria is almost $0$ and the third one almost $1$, then the system suddenly jumps to another configuration where two other species are almost extinguished and so on in a cyclic way~\cite{MayLeonard}. Introducing the mutation, one can prove~\cite{Mobilia1} that the coexistence equilibrium can become stable if the mutation rate is large enough, while if the mutation rate decreases the system undergo through a Hopf bifurcation and a limit cycle is created. The 2D spatial extension of the model is characterised by a coexistence of species and by the development of spatio-temporal patterns, more precisely, spiral waves~\cite{Mobilia, Mobilia1}.

Let us now introduce the 1D individual based description of the above presented model. The three species of bacteria evolve on a linear chain composed by $\Omega$ cells with periodic boundary conditions. Each cell has a finite carrying capacity, say $N$ and hereby assumed to be the same for all the cells. The number of bacteria of species $A$, respectively $B$ and $C$, in the $j$--th cell, will be denoted by $A^j$, respectively $B^j$ and $C^j$. Because of the above assumption we also have to consider the effect of excluded volume, denoting by $E^j$  the number of empty spaces available in the cell $j$, we finally got:
\begin{equation}
\label{eq:spacecontr}
A^j+B^j+C^j+E^j=N\quad \forall \,j \, .
\end{equation}

To simplify the notations we will rename the species as follows $A^j \rightarrow S_1^j$, $B^j \rightarrow S_2^j$ and $C^j \rightarrow S_3^j$, where the index $j$ represents the cell, that is the space.

A bacterium can move from one cell to one of its two neighbouring ones if enough space is available~\footnote{We could also have consider the possibility for two bacteria to hop, that is exchange their places in two neighbouring cells. Because this new action will not have introduced any new phenomenon, we decided to not consider it and to have a model as simple as possible.}, i.e. the number of vacancies in the destination cell is strictly positive, assuming all bacteria to have the same diffusivity coefficient, hereby named $\delta$, we obtain:
\begin{eqnarray}
\label{eq:move}
S_i^j +E^k&\autorightarrow{$\delta$}{}& S_{i}^k+E^j, \quad \forall \,j  \,\text{and}\, k\in j\, ,\notag\\
\end{eqnarray}
where we introduced the notation $k\in j$ to denote that $k$ is any of the neighbouring cell of the $j$--th cell.

A bacterium can reproduce and the offspring will occupy an available space in the same cell, if enough space is at its disposal. We will assume all the bacteria to have the same reproductivity coefficient hereby named $\beta$:
\begin{eqnarray}
\label{eq:reprodii}
S_i^j+E^j &\autorightarrow{$\beta$}{}& S_i^j+S_i^j \quad \forall \, j\, .
\end{eqnarray}

In the spirit of the cyclic interaction of the RPS model, we assume that bacteria of species $S_i$ dominates over $S_{i+1}$ while being dominated by $S_{i-1}$, where we define $S_{3+1}\equiv S_1$ and $S_{0}\equiv S_{3}$. To simplify we further assume that the competition rate is the same for all the bacteria and it will be denoted by $\sigma$. Assuming such interactions to hold only among bacteria living in the same cell, we get:
\begin{eqnarray}
\label{eq:selii}
S_i^j+S_{i+1}^{j} &\autorightarrow{$\sigma$}{}&S_{i}^j+E^j\quad \forall \, j\, .
\end{eqnarray}

Let us also consider the presence of a process of dominance-replacement, with rate $\zeta$:
\begin{eqnarray}
\label{eq:selii1}
S_{i}^{j}+S_{i+1}^j &\autorightarrow{$\zeta$}{}& S_i^j+S_i^j\quad \forall \, j\, .
\end{eqnarray}

Finally the mutation introduces the possibility that a bacterium of one species can transform into one of the other species, that is :
\begin{eqnarray}
\label{eq:seliia}
S_i^j \autorightarrow{$\mu$}{} S_{i+1}^j \quad \text{and} \quad S_i^j \autorightarrow{$\mu$}{} S_{i-1}^j\quad \forall \, j \, .
\end{eqnarray}

The time evolution of the above model is completely described by the Master Equation, governing the evolution of the probability to have, at any given time, an amount of bacteria $A$, $B$ and $C$ in any cells. As already stated, the van Kampen expansion will provide, at first order, the mean field description of the system, we thus decide to postpone a detailed analysis of the dynamics of such model to Section~\ref{ssec:timeevolav}, after having introduced the Master Equation.

\section{The Master Equation and the van Kampen expansion}
\label{sec:ME}

The state of the system at any time $t$ is completely determined by the amount a bacteria of each species in each cell, thus because of the constraint~\eqref{eq:spacecontr}, it will be enough to have : ${\mathbf{n}}(t)=[A^1(t),B^1(t),C^1(t), \dots,A^\Omega(t),B^\Omega(t),C^\Omega(t)]$. The goal of this section is to introduce a framework where the system evolution can be studied, that is the so called (chemical) Master Equation. 

Starting from the chemical reactions~\eqref{eq:move}~--~\eqref{eq:seliia} it is possible to compute the transition probabilities, $T\left(\mathbf{n}^\prime|\mathbf{n}\right)$, i.e. the probability for the system to jump from state ${\mathbf{n}}(t)$ to a new compatible one ${\mathbf{n}}^{\prime}(t^{\prime})$, in small time interval, $\lvert t-t^\prime\rvert<<1$. The Master Equation is thus obtained by taking into account all the possible ways the system can leave a given state, $\mathbf{n}$, and reach a new state $\mathbf{n}^\prime$:
\begin{equation}
 \label{eq:ME}
\frac{dP}{dt}(\mathbf{n},t)=\sum_{\mathbf{n}^\prime\neq \mathbf{n}}\left[T(\mathbf{n}|\mathbf{n}^\prime)P(\mathbf{n}^\prime,t)-T(\mathbf{n}^\prime|\mathbf{n})P(\mathbf{n},t)\right]\, .
 \end{equation}
 
More precisely the probability that a bacterium $S_i^j$, $i\in\{1,2,3\}$, moves from the $j$--th cell to the $k$--th one, is given by~\footnote{To lighten the notations, we hereby indicate only the variables whose values change because of the transition.}:
\begin{equation}
 \label{eq:Texch}
 T\left(S_i^j+1,S_i^k-1,E^j-1,E^k+1|S_i^j,S_i^k,E^j,E^k\right)=\frac{\delta}{z\Omega}\frac{S_i^jE^j}{N^2}\quad \text{where $k\in j$}\, .
 \end{equation}
The factor $z$ stands for the number of nearest neighbours cells, being all the movements equally probable, in the following 1D case with nearest neighbours we will set $z = 2$. Let us observe that the above formula is based on the assumption that the transition probabilities are proportional to the concentration of each species involved in the reaction and to the rate of success of the reaction, that is we assume that in each cell the bacteria are well stirred.

The reproduction process~\eqref{eq:reprodii} of one $S_i$ bacterium in the $j$--th cell corresponds to the transition probability:
\begin{equation}
 \label{eq:Trii}
 T\left(S_i^j+1,E^j-1|S_i^j,E^j\right)=\frac{\beta}{\Omega}\frac{S_i^jE^j}{N^2}\, ,
 \end{equation}
whereas to the selection mechanism~\eqref{eq:selii}, where a $S_{i}^j$ bacterium in the  $j$--th cell, fights against and destroys an $S_{i+1}^j$ bacterium we associate:
\begin{equation}
 \label{eq:Tsii}
 T\left(S_{i+1}^j-1,E^j+1|S_i^j,S_{i+1}^j\right)=\frac{\sigma}{\Omega}\frac{S_{i+1}^jS_i^j}{N^2}\, .
 \end{equation}
 The remaining cases, Eq.~\eqref{eq:selii1} and~\eqref{eq:seliia}, can be handled similarly. Let us however observe the different normalisation for the mutation, being a \lq\lq mono--molecular\rq\rq reaction:
 \begin{equation}
 \label{eq:mu}
 T\left(S_{i\pm1}^j+1|S_i^j\right)=\frac{\mu}{\Omega}\frac{S_{i}^j}{N}\, .
 \end{equation}

To simplify the notations and to prepare the set up for the following van Kampen expansion, we introduce the step operator:
\begin{equation}
\epsilon_{ij}^{\pm}f(\dots,S_i^j,\dots)=f(\dots,S_i^j\pm 1,\dots),
\end{equation} 
where $f$ represents a generic function, the index $i$ denotes the different species and $j$ the spatial location. In this way we can rewrite the Master Equation~\eqref{eq:ME} as follows:
\begin{eqnarray}
 \label{eq:ME1}
\frac{dP}{dt}(\mathbf{n},t)&=&\sum_{i=1}^3\sum_{j=1}^{\Omega}\Big\{ \sum_{k\in j}\Big[(\epsilon_{i\,j}^+\epsilon_{i\,k}^{-}-1) T(S_i^j-1,S_i^k+1,E^j+1,E^k-1|S_i^j,S_i^k,E^j,E^k)\notag\\
&+&(\epsilon_{i\,j}^-\epsilon_{i\,k}^{+}-1) T(S_i^j+1,S_i^k-1,E^j-1,E^k+1|S_i^j,S_i^k,E^j,E^k)\Big]\Big\}P(\mathbf{n},t)\notag\\
&+&\sum_{i=1}^3\sum_{j=1}^{\Omega}\Big[(\epsilon_{ij}^+\epsilon_{i+1\,j}^{-}-1) T(S_i^j-1,S_{i+1}^j+1,\dots | S_i^j,S_{i+1}^j\dots)\notag\\
&+&(\epsilon_{i\,j}^+\epsilon_{i-1\,j}^{-}-1) T(S_i^j-1,S_{i-1}^j+1,\dots |S_i^j,S_{i-1}^j,\dots)\notag\\
&+&(\epsilon_{i\,j}^--1) T(S_i^j+1,\dots | S_i^j,\dots)+(\epsilon_{i+1\,j}^+-1) T(S_{i+1}^j-1,\dots |S_{i+1}^j,\dots)\notag\\
&+&(\epsilon_{i+1\,j}^+\epsilon_{i\,j}^{-}-1) T(S_{i+1}^j-1,S_{i}^j+1,\dots |S_{i+1}^j, S_i^j,\dots)
\Big]P(\mathbf{n},t).
\end{eqnarray}

Such equation is difficult to handle analytically and one has to resort to approximate techniques to progress in the study, a possibility is to use the celebrated van Kampen system size expansion~\cite{vanKampen}, a perturbative calculation that recovers the mean-field system at the first order and a Fokker-Planck equation describing the fluctuations, at the second order.

The starting point is the following ansatz, the number of bacteria in each cell is given by a \lq\lq regular\rq\rq  function plus a stochastic contribution, vanishing in the limit of large system size:
$$\frac{S_i^j}{N}=\phi_i^j+\frac{\xi_i^j}{\sqrt{N}}\, ,$$
more precisely, $\phi_i^j$ will denote the deterministic concentration, in the limit $N\rightarrow \infty$, of the species $i$ in cell $j$, while $\xi_i^j$ is a stochastic variable that quantifies the intrinsic fluctuation that perturbs the idealised mean field deterministic solution $\phi_i^j$. The amplitude factor $1/\sqrt{N}$ encodes the finite size of the system and it is the small parameter in the following perturbative analysis. 

Putting the van Kampen ansatz into the master equation, developing the step operators, collecting together the terms with the same power of $\sqrt{N}$ and rescaling time by ${t}/{(N\Omega)}$, one recovers at the first order:
\begin{eqnarray}
 \begin{cases}
\frac{d\phi_1^j}{dt}(t)&=\phi_1^j[\beta(1-r^j)-\sigma \phi_3^j+ \zeta (\phi_2^j-\phi_3^j)]+\mu(\phi_3^j+\phi_2^j-2\phi_1^j)+\delta\Delta \phi_1^j+\delta(\phi_1^j\Delta r^j-r^j \Delta \phi_1^j)\\
\frac{d\phi_2^j}{dt}(t)&=\phi_2^j[\beta(1-r^j)-\sigma \phi_1^j+ \zeta (\phi_3^j-\phi_1^j)]+\mu(\phi_3^j+\phi_1^j-2\phi_2^j)+\delta\Delta \phi_2^j+\delta(\phi_2^j\Delta r^j-r^j \Delta \phi_2^j)\\
\frac{d\phi_3^j}{dt}(t)&=\phi_3^j[\beta(1-r_j)-\sigma \phi_2^j+ \zeta (\phi_1^j-\phi_2^j)]+\mu(\phi_1^j+\phi_2^j-2\phi_3^j)+\delta\Delta \phi_3^j+\delta(\phi_3^j\Delta r^j-r^j \Delta \phi_3^j),
 \end{cases}\label{eq:PDE}
 \end{eqnarray}
where we introduced $r^j=\sum_{i=1}^{3} \phi_i^j$ and the discrete Laplacian $\Delta f_j:=\frac{2}{z}\sum_{k\in j}\left(f_k-f_j\right)$. Let us remember that in the present case of 1D system with nearest neighbours $z=2$ and $k \in \{i-1,i+1\}$.

The effect of the finite carrying capacity reflects in the above mean--field equations through the non--linear cross diffusion terms $(−\phi^j_i\Delta r^j -r^j \Delta \phi^j_i)$ which appear to modify the conventional Fickean behaviour. These are second order contributions in the concentrations and are therefore important in the regime of high densities \cite{fanellimckane, Fanelli_2011, cianci_DNC}.

The expansion to the next leading order will determine a Fokker-Planck equation describing the probability distribution of the fluctuation, $\Pi(\mathbf{\xi},t)$, that will be introduced and analysed in Section~\ref{S3}.

\section{Analysis of the Mean-field system}
\label{ssec:timeevolav}

Let us start our analysis by considering the spatially homogeneous solutions of the previous system~\eqref{eq:PDE}, namely we assume the following limit does exist and it is independent from the spatial index $j$:
\begin{equation}
 \label{eq:meanfield}
\lim_{N\rightarrow \infty}\frac{S_i^j}{N}=\phi_i\, ,
\end{equation}
hence~\eqref{eq:PDE} rewrites:
\begin{eqnarray}
 \label{eq:ODEmfsh}
 \frac{d\phi_i}{dt}(t)&=\phi_i[\beta(1-(\phi_i+\phi_{i+1}+\phi_{i-1}))-\sigma \phi_{i-1}+ \zeta (\phi_{i+1}-\phi_{i-1})]+\mu(\phi_{i-1}+\phi_{i+1}-2\phi_{i})\quad i\in\{1,2,3\}\, , \end{eqnarray}
where we used once again the notation $\phi_{3+1}\equiv \phi_1$ and $\phi_{0}\equiv \phi_3$.

A straightforward analysis~\cite{Mobilia} of the above system shows that it admits the equilibrium point:
\begin{equation}
 \label{eq:equil}
 S^*= \left(\frac{\beta}{3\beta+\sigma},\frac{\beta}{3\beta+\sigma},\frac{\beta}{3\beta+\sigma}\right)\, ,
\end{equation}
and the system behaviour can be summarised by:
\begin{itemize}
  \item for $\mu>\mu_H$, where $\mu_H=\frac{\beta\sigma}{6(3\beta+\sigma)}$, $S^*$ is stable focus and the trajectories generically converge to $S^*$;
\item $\mu=\mu_H$ there is a supercritical Hopf bifurcation and the associated frequency is $\omega_H=\frac{\sqrt{3}\beta (\sigma+2\zeta)}{2(3\beta+\sigma)}$;
\item $\mu<\mu_H$, $S^*$ is an unstable focus and a stable limit cycle emerges from the Hopf bifurcation.
\end{itemize}
Because of the cyclic competition, the Jacobian matrix of the system~\eqref{eq:ODEmfsh} is a circulating matrix, that evaluated at $S^*$ reduces to:
\begin{equation}
 \label{jacobian1}
 J_{S^*}=\left(
\begin{matrix}
a_0 & a_2 & a_1\\
a_1 & a_0 & a_2\\
a_2 & a_1 & a_0
\end{matrix}
\right)\, ,
\end{equation}
 where:
\begin{eqnarray}
&a_0=\frac{-\beta^2-2\mu(3\beta+\sigma)}{3\beta+\sigma}\quad
a_1=\frac{1}{3\beta+\sigma}\left[-\beta^2-\sigma\beta+\mu(3\beta+\sigma)-\delta\beta\right]\quad
a_2=\frac{1}{3\beta+\sigma}\left[-\beta^2+\zeta\beta+\mu(3\beta+\sigma)\right]\, .
\end{eqnarray}  
The eigenvalue are thus easily obtained:
\begin{eqnarray}
&\lambda_0=-\beta,\notag\\
&\lambda_{1,2}=\frac{1}{2(3\beta+\sigma)}\left[\left(-6\mu(3\beta+\sigma)+\sigma\beta\right)\pm i\sqrt{3}\left(\sigma\beta+2\zeta\beta\right)\right].
\end{eqnarray} 

\begin{figure}[htbp!]
\centering
\includegraphics[width=8cm]{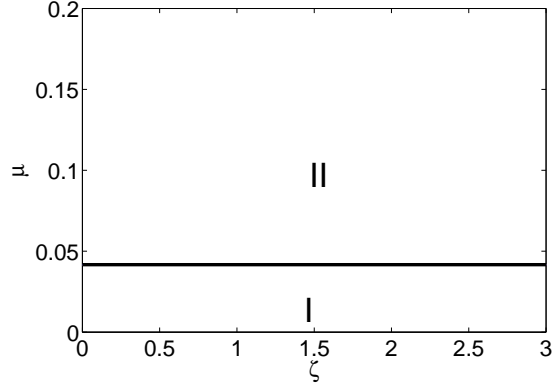}
\caption{Parameters plane $(\zeta,\mu)$. We fix $\beta=1$ and $\sigma=1$ and we delimited two zones corresponding to two different dynamical behaviours: (zone I) existence of a stable limit cycle, (zone II) presence of a stable fixed point. The zones are separated by the line $\mu=\mu_H$ where the system undergoes through a Hopf bifurcation.}
\label{figura1}
\end{figure}

In Figure~\ref{figura1} we summarise the dynamics of the homogeneous system as a function of two parameters $\mu$ and $\zeta$ once we fixed the remaining two $\beta$ and $\sigma$. Let us observe that the plane $(\zeta,\mu)$ is divided in two zones, in the first one (zone I) the system presents a stable limit cycle and an unstable fixed point, while in the second one (zone II) there is a stable fixed point. The line separating the two zones, $\mu=\mu_H$, corresponds to the supercritical Hopf bifurcation.

We are now able to recover the space dependence and interested in identifying conditions yielding to a spontaneous amplification of the perturbation and eventually translate in the emergence of stochastic patterns. To this end, and following the standard approach, we consider the linear stability analysis of the full system~\eqref{eq:PDE} close to the homogeneous solution $\phi_i=S^*$ for $i=1,2$ and $3$. To better understand the system's behaviour we analyse the linearised system in the Fourier space, where the Jacobian of the non-homogeneous system, $J_{NH}^*$, reads:
$$J_{NH}^*=J_{S^*}+D^*\tilde{\Delta}\, ,$$
$\tilde{\Delta}$ is the Fourier transform of the Laplacian and $D^*$ is the diffusion matrix evaluated at $S^*$:
\begin{equation}
 \label{DDjacobian1}
 D^*=\delta\left(
\begin{matrix}
b_0 & b_1 & b_1\\
b_1 & b_0 & b_1\\
b_1 & b_1 & b_0
\end{matrix}
\right)\, ,
\end{equation}
where:
\begin{equation}
b_0=\beta+\sigma/(3\beta+\sigma)\quad\text{and}\quad
b_1=\beta/(3\beta+\sigma)\, .
\end{equation}

The eigenvalues of the linearized system, in Fourier space, are (once we approximate $\tilde{\Delta}$ with $-k^2$): 
\begin{eqnarray}
&\rho_0=-\beta-\delta k^2 \, ,\\
&\rho_{1,2}=\frac{1}{2(3\beta+\sigma)}\left[-6\mu(3\beta+\sigma)+\sigma\beta-2\delta k^2(2\beta+\sigma)\right]\pm\frac{i\sqrt{3}}{2(3\beta+\sigma)}\left[\sigma\beta+2\zeta\beta\right]\, .
\end{eqnarray}
We can observe that $\rho_0$ is always negative and thus corresponds to a stable direction also for the spatial system. 
The interesting dynamics is hence reduced to study the other two eigenvalues. The imaginary parts of these eigenvalues don't depend on $k^2$ and thus they are the same as the homogeneous case. The real parts differ for the new term $-2\delta k^{2}(2\beta+\sigma)$. So we can conclude that the mode $k=0$ has the same behaviour as the aspatial system; moreover because the mode $k=0$ dominates the dynamics induced by the other modes, even if the real parts of $\rho_{1,2}$ can change their signs with respect to the aspatial case, they cannot introduce any new dynamical behaviour. So the patterns we eventually find would not due to a Turing like mechanism because no eigenvalue will change its real part and thus destabilise the homogeneous solution, they will be purely stochastic ones.

\begin{figure}[htbp!]
%[][tb]
\begin{tabular}{cc}
\includegraphics[width=8cm]{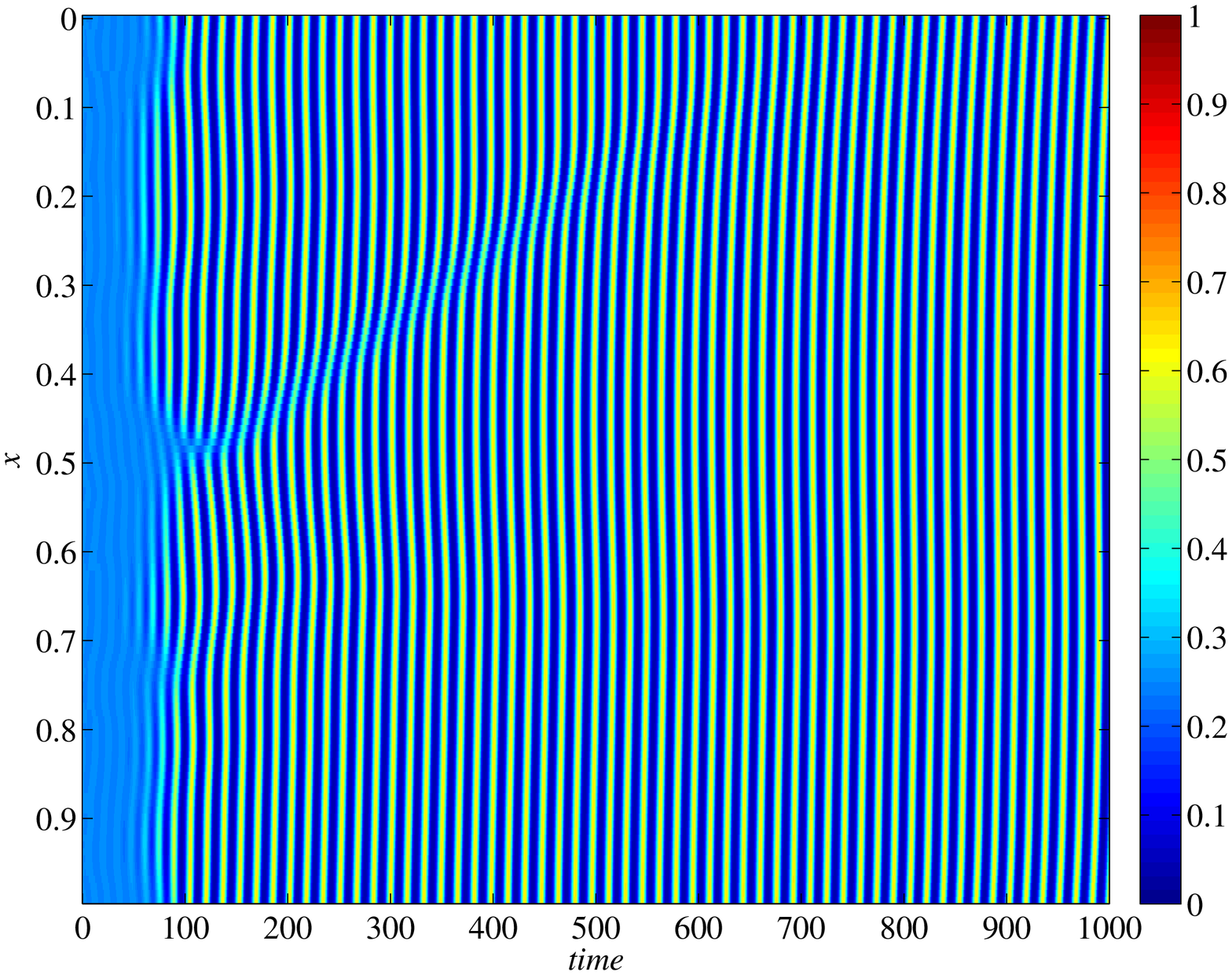}&
\includegraphics[width=8cm]{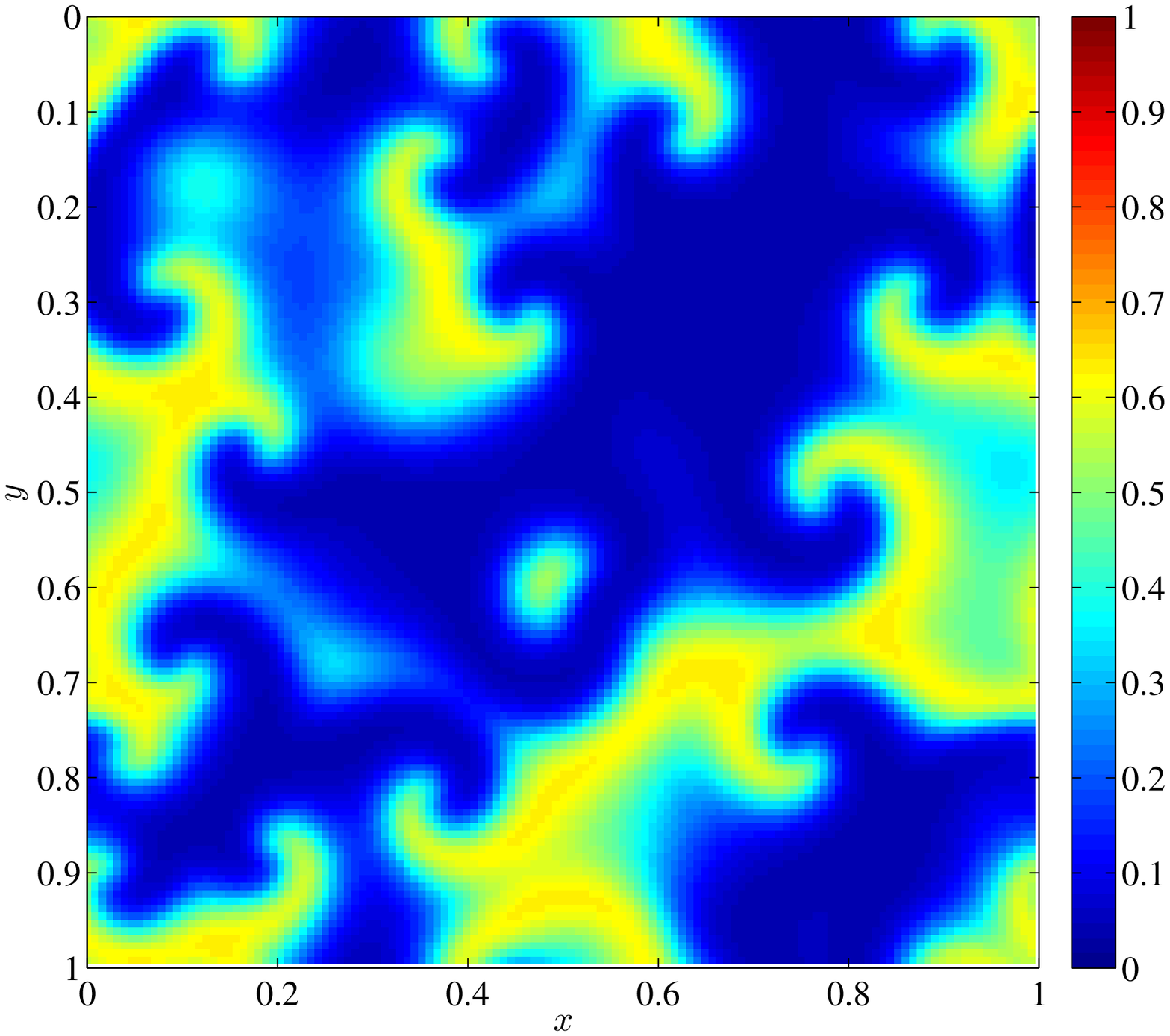}\\
(a) & (b)\\
\end{tabular}
%\caption{Deterministic patterns. Panel (a): numerical integration of 1D system, we report the time evolution of the concentrations of $A$ (red), $B$ (green) and $C$ (blue). Panel (b): generic time snapshot of the spiral waves for the 2D model, we used the same colour code as in panel (a). Both simulations refer to the following parameters values: $\zeta=0.6$, $\beta=1$, $\sigma=1$, $\delta=1/128^2$ and $\mu=0.02<\mu_H=1/24$.}
\caption{Deterministic patterns. Panel (a): numerical integration of 1D system, we report the time evolution of the concentrations of species $A$, being the ones for $B$ and $C$ similar because of the cyclic dominance. Panel (b): generic time snapshot of the spiral waves for the 2D model, still for species $A$. Both simulations refer to the following parameters values: $\zeta=0.6$, $\beta=1$, $\sigma=1$, $\mu=0.02<\mu_H=1/24$, $\delta=1/L^2$ and the spatial domain has been discretised into $L=128$ identical cells.}
 \label{pattern}
\end{figure}

The solutions of the system~\eqref{eq:PDE} for parameters in zone II converge to the spatial homogeneous solution, on the other hand once parameters are fixed in the zone I one can obtain stable patterns, i.e. spatially organised and time synchronised structures, as reported in Figure~\ref{pattern}, where we report on the left panel the results of a numerical integration of the 1D system and on the right panel we report a snapshot of the numerical integration of the 2D model where spiral waves can be observed as already reported by~\cite{Mobilia}.

\begin{figure}[htbp!]
%[][tb]
\includegraphics[width=8cm]{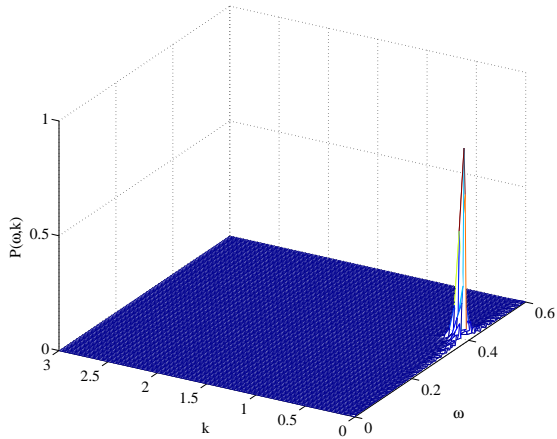}
\caption{Power spectrum in the spatio--temporal domain for the 1D model for the same parameters used for Fig.~\ref{pattern}.} 
\label{fig:1DXTFFT}
\end{figure}

\begin{figure}[htbp!]
%[][tb]
\begin{tabular}{cc}
\includegraphics[width=8cm]{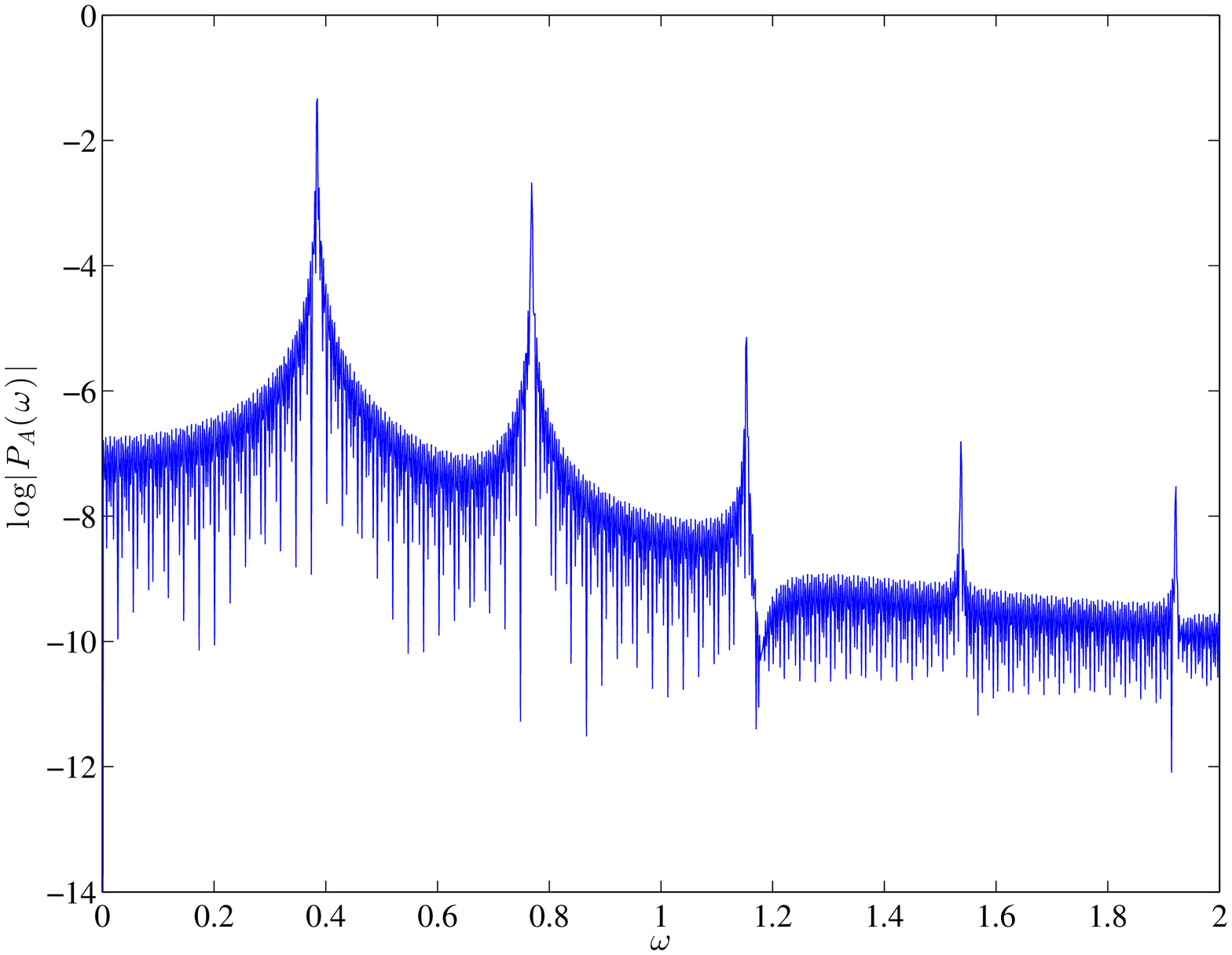}&
\includegraphics[width=8cm]{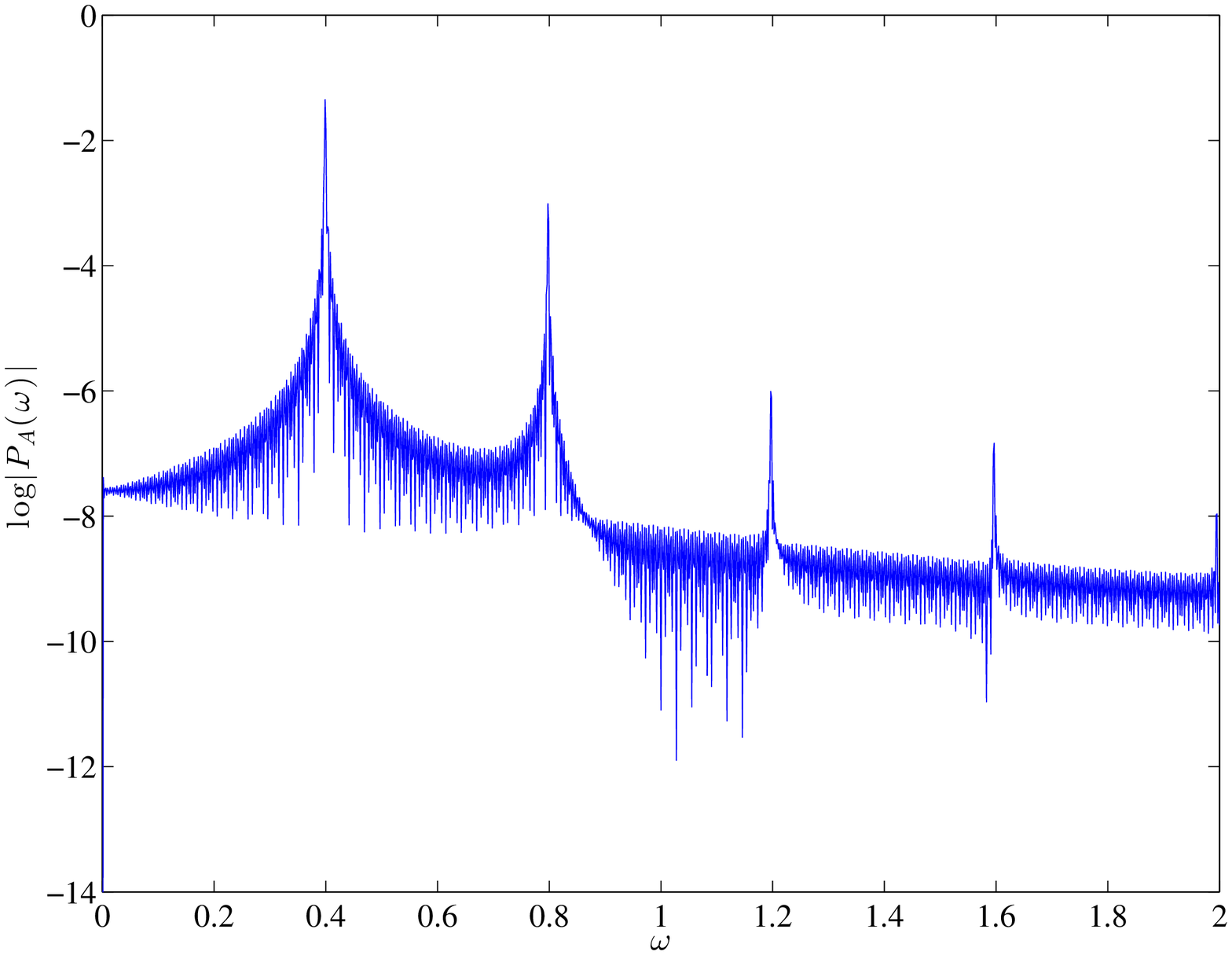}\\
(a) & (b)\\
\end{tabular}
\caption{Power spectrum in the temporal domain. Panel (a) represents the logarithm of the power spectrum for the 1D model for a generic spatial position. Panel (b) represents the logarithm of the power spectrum of the 2D model for a fixed generic spatial coordinates. Simulations have been done using the same set of parameters used for Fig.~\ref{pattern}.} 
\label{patternspettri}
\end{figure}

A more complete understanding of the patterns presented in Fig.~\ref{pattern} can be obtained by analysing the Fourier spectrum, both spatial and temporal one, see Figure~\ref{fig:1DXTFFT}. Moreover in Figure~\ref{patternspettri} we plot the power spectrum of the temporal Fourier transform of the patterns shown in Figure~\ref{pattern}, we can observe that in both 1D and 2D cases, the spectra behave in a similar way with a clear peak at a frequency that is close to the Hopf frequency; also the spatial ones (data not shown) exhibit a similar behaviour with a decrease of the spectrum as a function of the spatial modes. The similarity of such behaviours allows us to term the pattern observed in the panel (a) of Figure \ref{pattern} {\em one dimensional spiral waves}.

In previous studies~\cite{Mobilia, Mobilia1} authors studied the robustness of the spiral waves against the noise and of the system parameters and concluded that stochastic spirals waves exist for parameters in zone I, that is in the same range as for the mean--field approximation. Our goal is to prove that stochastic spiral waves do exist in a larger parameters domain, covering part of the zone II, where the mean--field solutions converge to the homogeneous one. To achieve our goal we will characterise such patterns by analysing the power spectrum of the fluctuations, to this end we need to introduce and study the Fokker--Planck equation that governs the evolution of the fluctuations. Let us observe that our approach is completely different from the Complex Ginzburg--Landau equation used~\cite{frey,Mobilia3} and the multi--scale method used in~\cite{Mobilia1}.

\section{Fokker-Planck equation and fluctuations power spectrum}
\label{S3}
As already stated, the next to the leading order in the van Kampen system size expansion allows to characterise the distribution of the fluctuations, $\Pi(\mathbf{\xi},t)$, in fact a cumbersome computation allows us to derive a Fokker--Planck equation for $\Pi(\mathbf{\xi},t)$:
\begin{equation}
\label{FP}
\frac{d\Pi}{dt}(\mathbf{\xi},t)=-\sum_{i=1}^3\sum_{j=1}^{\Omega}\frac{\partial}{\partial\xi_i^j}\Big[Q_{i}^{j}(\xi)\Pi(\xi,t)\Big]+\frac{1}{2}\sum_{i=1}^3\sum_{j=1}^{\Omega}\sum_{p\in j}\sum_{h=1}^3 R_{ih}\frac{\partial^2\Pi(\xi,t)}{\partial\xi_i^j\partial\xi_h^p}\, ,
\end{equation}
where we denoted by $\mathbf{\xi}=(\xi_1^1,\xi_2^1,\xi_3^1,\ldots,\xi_1^\Omega,\xi_2^\Omega,\xi_3^\Omega)$.
In the above expression $Q_i^j$ can be expressed in terms of the Jacobian $M$ of the mean--field system~\eqref{eq:ODEmfsh} evaluated at the fixed point $S^*$. For a sake of clarity we split $M$ into two parts: the one named $M^{(d)}$, depending on the diffusion part of the mean--field equation, i.e. involving the Laplacian, and the remaining one, $M^{(r)}$, i.e. associated to the reaction terms. More explicitly we get:
\begin{eqnarray}
M_{i\, i-1}^{(r)}&=&\mu-(\beta +\sigma +\zeta) S^*\notag\\
M_{i\, i+1}^{(r)}&=&\mu-(\beta +\zeta) S^*\notag\\
M_{ii}^{(r)}&=&-2\mu+\beta\left(1-4S^*\right)-\sigma S^*\\\notag%+\delta\left(S^*-S^*\right)\notag\\
M_{ih}^{(r)}&=&0 \quad\text{in all the remaining cases}\, .
\end{eqnarray}
On the other hand the spatial contribution is:
\begin{eqnarray}
M_{ii}^{(d)}&=&\delta[1+2S^*]\notag\\
M_{ih}^{(d)}&=&\delta S^* \quad \text{for all $h \neq i$}\, .
\end{eqnarray}
Finally:
\begin{equation}
M_{ih}=M_{ih}^{(r)}+M_{ih}^{(d)}.
\end{equation}
We can thus write:
\begin{equation}
\label{eq:QM}
Q_{i}^{j}(\xi)=\sum_{h=1}^{3} \left[M_{ih}^{(r)}\xi_h^j+ M_{ih}^{(d)}\Delta\xi_h^j\right]\, ,
\end{equation}
where $\Delta$ is the discrete Laplacian.

The same splitting can be applied to the matrix $R$, still evaluated at the fixed point $S^*$:
\begin{eqnarray}
R_{ii}^{(r)}&=&\beta S^*(1-3 S^*)+4\mu S^*+2\zeta S^*+\sigma (S^*)^2\notag\\
R_{i\, i-1}^{(r)}&=&-\mu S^*\notag\\
R_{i\, i+1}^{(r)}&=&-\mu S^*+\zeta (S^*)^2\notag\\
R_{ih}^{(r)}&=&0 \quad\text{in all the remaining cases}\, ,
\end{eqnarray}
and
 \begin{eqnarray}
R_{ii}^{(d)}&=&-2 \delta S^*[1-3 S^*]\notag\\
R_{ih}^{(d)}&=&0 \quad h \neq i\, ,
\end{eqnarray}
and thus we get:
\begin{equation}\label{R}
R_{ih}^{jp}=\left[R_{ih}^{(r)}+ R_{ih}^{(d)}\Delta_{jp}\right]\, , 
\end{equation}
where $\Delta_{jp}$ denotes the discrete Laplacian associated to the linear lattice with periodic boundary conditions :
\begin{equation*}
\Delta_{jp}=W_{jp}-2\delta_{jp}\, ,
\end{equation*}
being $\delta_{jp}$ the Kronecker delta and $W_{jp}$ the matrix given by:
\begin{equation*}
\begin{cases}
1 & \text{if \quad $j=p\pm 1$}\\
0 & \text{otherwise}\, .
\end{cases}
\end{equation*}

%The interested reader can find the detailed computations needed to get the previous expressions in the appendix.
To handle the Fokker--Planck is not so straightforward we thus prefer pass to an equivalent Langevin equation \cite{Gardiner1985}, where the noise term intrinsically depends on the system fluctuations.
\begin{equation}
\label{eq:xidot}
\frac{d \xi_i^j}{dt} = Q_{i}^{j}(\xi)+\eta_i^{j}(t)\, ,
\end{equation}
where the stochastic contribution satisfies the following relations:
\begin{equation}
<\eta_i^{j}(t)>=0\quad\text{and}\quad <\eta_i^{j}(t)\eta_h^{p}(t')>=R_{ih}^{jp}\delta(t-t')\, .
\end{equation}
Recalling the~\eqref{eq:QM}, introducing spatial and temporal Fourier variables Eq.~\eqref{eq:xidot} reads (denoting by $\tilde{f}^{\mathbf{k}}(\omega)$ the spatio-temporal Fourier transformation of the function $f(x,t)$): 
\begin{equation}
-i\omega\tilde{\xi_i}^{\mathbf{k}}(\omega) = \sum_{h=1}^{3}  [\tilde{M}_{ih}^{r} \tilde{\xi}_h^{\mathbf{k}}+\tilde{M}_{ih}^{d} \tilde{\Delta\xi}_h^{\mathbf{k}}]+ \tilde{\eta}_i^{\mathbf{k}}(\omega)\, ,
\label{eq:langevin}
\end{equation}
where, 
\begin{equation}\label{eq:lambda}
<\tilde{\eta}_i^{\mathbf{k}}(\omega)\tilde{\eta}_h^{\mathbf{k'}}(\omega')> =\Omega \tilde{R}_{ih}^{\mathbf{k}}\delta_{k,-k'}\delta_{\omega,-\omega'}\, .
\end{equation}

Eq~\eqref{eq:langevin} can be rewritten as:
\begin{equation}
\label{passaggio}
\sum_{h} (-i\omega \delta_{ih}-[\tilde{M}_{ih}^{r} \tilde{\xi}_h^{\mathbf{k}}+\tilde{M}_{ih}^{d} \tilde{\Delta\xi}_h^{\mathbf{k}}]) \tilde{\xi}_h^{\mathbf{k}}=\tilde{\eta}_i^{\mathbf{k}}(\omega)\, ,
\end{equation}
whose solution is:
\begin{equation}
\hat{\xi}_h^{\mathbf{k}}=\sum_{i} \left[\Phi_{ih}^{\mathbf{k}}(\omega)\right]^{-1}\hat{\eta}_i^{\mathbf{k}}(\omega)\, ,
\end{equation}
being
\begin{equation}
\Phi_{ih}^{\mathbf{k}}(\omega)=(-i\omega \delta_{ih}- [\tilde{M}_{ih}^{r} \tilde{\xi}_h^{\mathbf{k}}+\tilde{M}_{ih}^{d} \tilde{\Delta\xi}_h^{\mathbf{k}}])\, .
\end{equation}

We are now able to analytically compute the power spectrum $P_i(\mathbf{k},\omega)$ of the fluctuations for each species $i=\{1,2,3\}$: 
\begin{equation}\label{powerformula}
P_i(\mathbf{k},\omega)=<|\xi_i^{\mathbf{k}}(\omega)|^2>=
\Omega \sum_{j=1}^3\sum_{u=1}^3 [\Phi^{\mathbf{k}}(\omega)]_{ij}^{-1} \tilde{R}_{ju}^{\mathbf{k}} [\Phi^{\mathbf{k}\dagger}(\omega)]_{ui}^{-1}\, ,
\end{equation}
where $\Phi^\dagger=\bar{\Phi}^T$. 

Using the above formula for the power spectrum of the fluctuations we are able to study the microscopic system for parameters values outside the regions of deterministic order, that is part of zone II, with the goal of looking for the signatures of a spatio-temporal organisation. Because the mean-field deterministic model will not display the same patterns, the latter should ultimately reflects the discreteness of the investigated stochastic model. To complement our analytical results, we will compare the power spectrum~\eqref{powerformula} with the numerical one obtained through a spatio--temporal FFT of the solutions of the microscopic system got using a Gillespie's algorithm.

\begin{figure}[htbp!]
%[][tb]
\begin{tabular}{cc}
\includegraphics[scale=0.45]{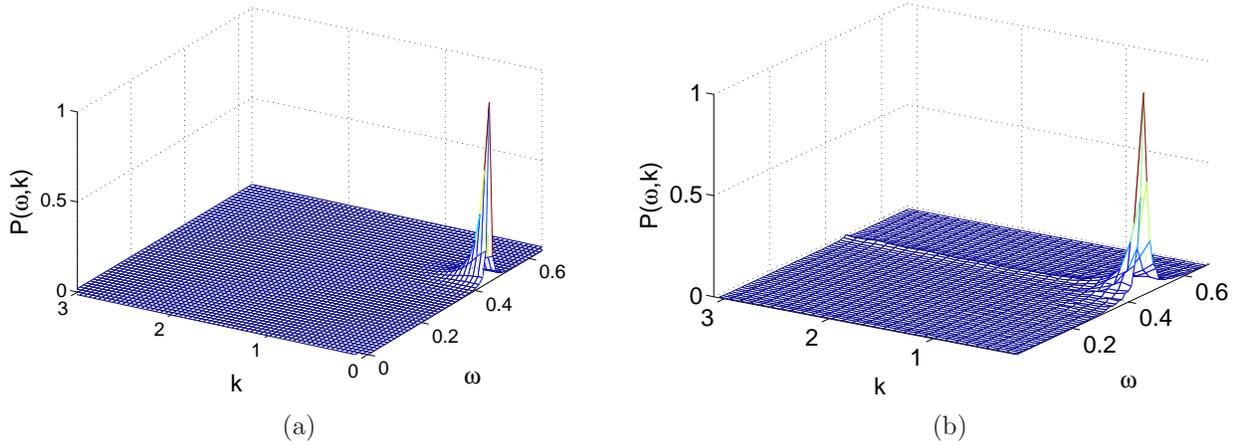}&
\includegraphics[scale=0.45]{powersimulazioni150medie.eps}\\
(a)&(b)
\end{tabular}
\caption{Power spectrum of the fluctuations. Panel (a): we report the analytical power spectrum given by formula~\eqref{powerformula}. Panel (b): the numerical power spectrum of the fluctuations for species $A$ is reproduced ($N=5000$  and $\Omega=32$). Both panels refer to parameters values in zone II; more precisely, we have set $\mu=0.043>\mu_H$, $\zeta=0.6$, $\beta=1$, $\sigma=1$ and $\delta=1/(2\cdot \Omega^2)$. The numerical power spectrum is obtained by averaging over $150$ independent realisations.}
 \label{power}
\end{figure}

Results reported in Figure~\ref{power} allow to conclude that the system has a spatio-temporal organisation also in (part of) zone II, in fact both power spectra present a clear peak in the $\omega$ variable, whose value is close to the Hopf frequency $\omega_H$, and rapidly decrease as $k^2$ increase. The agreement between the two spectra, gives a confirmation a posteriori of the validity of our analytical formula and the assumptions so far used. Let us observe that such spectra are very similar to the one reported in Figure~\ref{fig:1DXTFFT} for parameters values in the zone I where the system exhibits deterministic spiral waves. We can thus conclude that the results presented~\cite{Mobilia} hold in a larger domain, for instance for $\mu_H<\mu<\mu_*$, for some positive $\mu_*$. To determine such value, beyond which the noise will completely destroy the patterns, is surely an interesting question to which we will devote a forthcoming analysis.

\begin{figure}[htbp!]
\includegraphics[width=8cm]{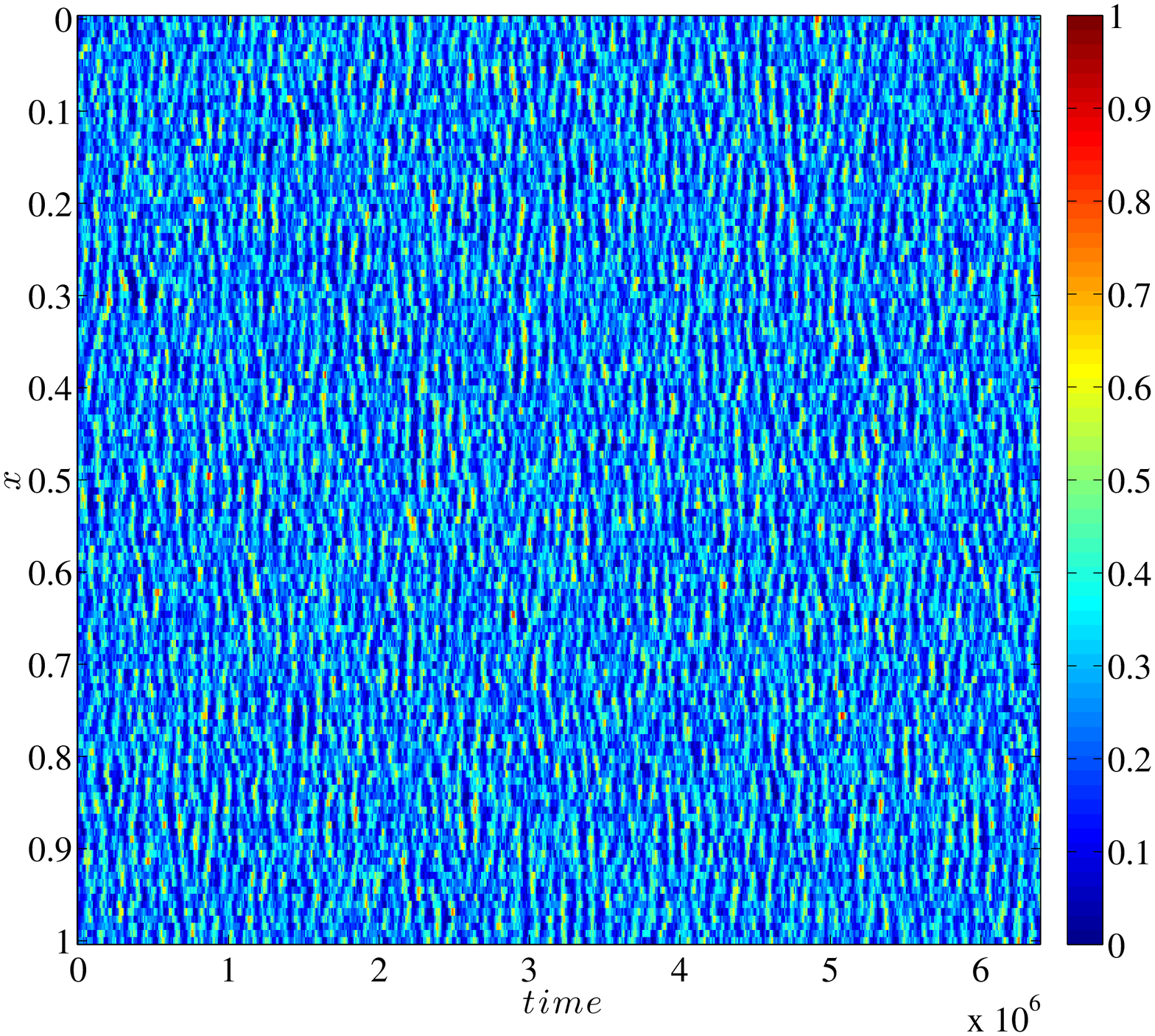}
\caption{Stochastic patterns. We report the time evolution of the concentration of $A$ obtained through a numerical integration of 1D system using the Gillespie algorithm to be compared with the one of Figure~\ref{pattern}. The parameters values are the same used to get the results presented in Figure~\ref{power} but $N=50$ and $\Omega=128$.
}
 \label{patternstocastici}
\end{figure}

In Figure~\ref{patternstocastici} we report a numerical simulation of the 1D individual based model, for the same parameters in zone II used to obtain the power spectrum presented in Figure~\ref{power}, where we can observe the spatio-temporal patterns, distorted by the noise. 

%In Figure~\ref{patternstocastici} we report a numerical simulation of the 1D and 2D systems, for the same parameters in zone II used to obtain the power spectrum presented in Figure~\ref{power}, where we can observe the spatio-temporal patterns, distorted by the noise.

\begin{figure}[htbp!]
\includegraphics[width=8cm]{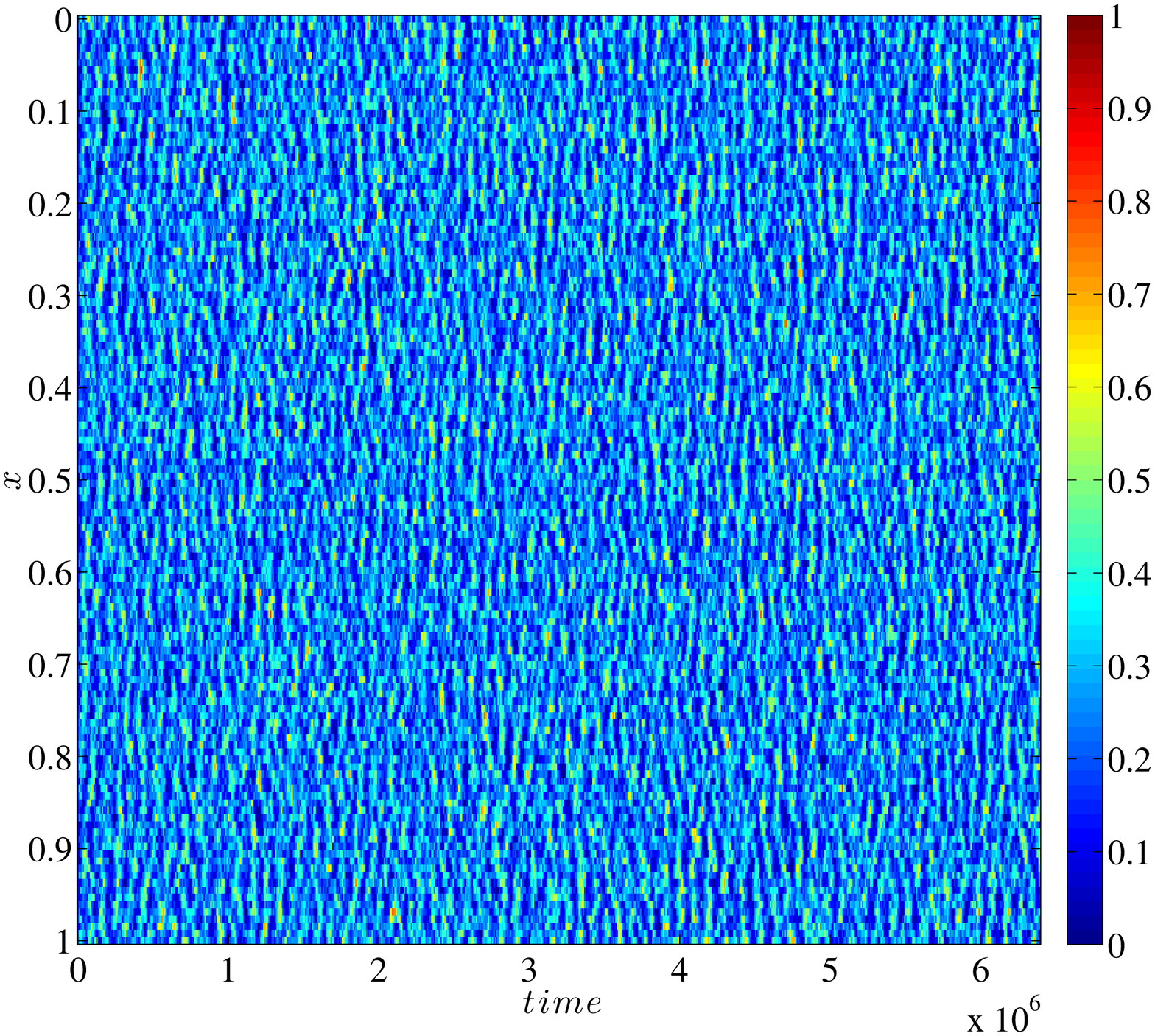}\quad
\includegraphics[width=9cm]{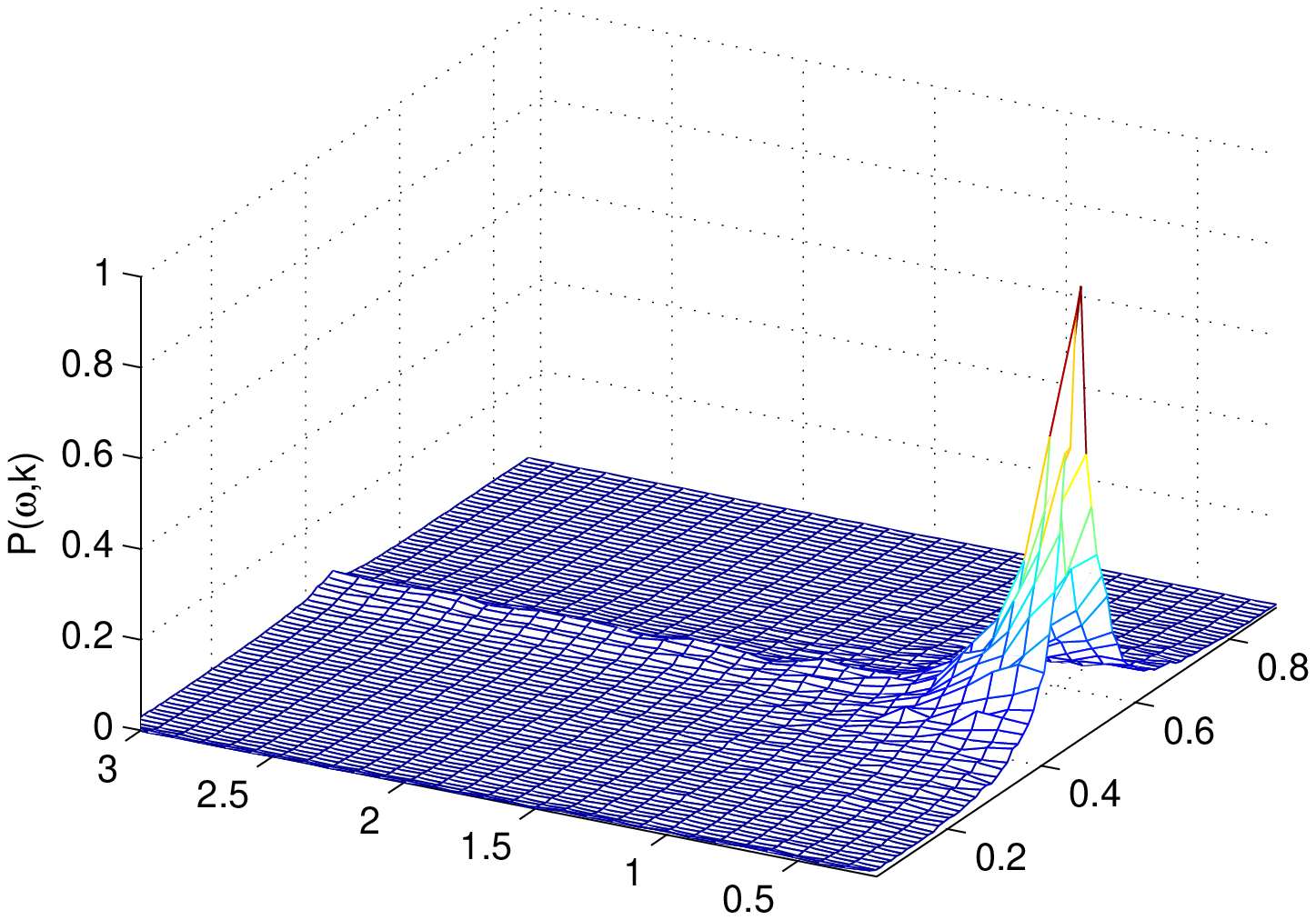}
\caption{Stochastic patterns for large $\mu$. Left panel, we report the time evolution of the concentration of $A$ obtained through a numerical integration of 1D system using the Gillespie algorithm with parameters $\mu=0.05$, $\zeta=0.6$, $\beta=1$, $\sigma=1$, $\delta=1/(2\Omega^2)$, $N=50$ and $\Omega=128$. Right panel, power spectrum in the spatio--temporal domain for the 1D model for the same parameters used in left panel. The numerical power spectrum is obtained by averaging over 150 independent realisations.} 
\label{fig:largemu}
\end{figure}

Let us finally conclude that such stochastic patterns do persist for large values of $\mu$ well beyond $\mu_H$ as can be numerically inspected in Fig.~\ref{fig:largemu}, where we report numerical simulations of the individual based model and the corresponding tempo--spatial Fourier transform for $\mu=0.05>>\mu_H$.

\section{Conclusion}

Spatio--temporal patterns are widely spread and encompass several research fields; in this scenario the Turing instability is one of the mechanisms that can be used to understand the emergence of such ordered patterns in reaction-diffusion models. Recent results have shown that such spatio--temporal patterns are robust agains intrinsic noise and they can persist for parameters values well beyond the ones fixed by the deterministic setting, e.g. the mean--field approximation.

In this paper we have considered a 1D version of the Rock--Paper--Scissor model with mutation~\cite{Mobilia1}, able to describe the coexistence of different species in a square lattice. We formulated the model as an individual based one, taking into account for the finite carrying capacity of each lattice cell. A preliminary analysis of the mean--field approximation allows us to recover the results by~\cite{Mobilia}, proving the existence of spatio--temporal \lq\lq wave--like spirals\rq\rq once the homogeneous model admits a limit cycle because of a Hopf bifurcation, $\mu<\mu_H$. 

However, being the proposed model inherently stochastic, we computed the Master Equation, describing the evolution of probability of being in a given state, and thus we performed a van Kampen system size expansion. Studying the first two terms of such approximation, we have been able to prove that such patterns are also present in (part of) the $\mu>\mu_H$ zone where the mean--field approximation predicts the existence of a a stable homogeneous solution; hence such structures are entirely driven by the intrinsic noise. As already stated our method is complementary to the ones used previously in the literature, the Complex Ginzburg--Landau equation and the multi--scale method.

\vspace{0.8cm}
{\it Acknowledgments}\\
The authors would like to warmly thank Duccio Fanelli for useful comments and discussions. \\
This research used computational resources of the \lq\lq Plateforme Technologique de Calcul Intensif (PTCI)\rq\rq located at the University of Namur, Belgium, which is supported by the F.R.S.-FNRS.\\
This paper presents research results of the Belgian Network DYSCO (Dynamical Systems, Control, and Optimization), funded by the Interuniversity Attraction Poles Programme, initiated by the Belgian State, Science Policy Office. The scientific responsibility rests with its author(s).

\bibliographystyle{unsrt}
\bibliography{bibliography}

\end{document}